\documentclass[a4paper,5p,twocolumn,11pt]{elsarticle}
\pdfoutput=1

\usepackage[T1]{fontenc}
\usepackage[utf8]{inputenc} 
\usepackage{lmodern}        

\usepackage{lettrine}
\usepackage{soul}
\usepackage{multirow}
\usepackage{graphicx}
\graphicspath{ {./images/} }

\usepackage{booktabs}
\usepackage{comment}
\usepackage{enumitem}

\usepackage{amsmath}
\usepackage[normalem]{ulem}

\usepackage{subcaption}
\usepackage{siunitx}

\usepackage[dvipsnames]{xcolor}

\definecolor{urlblue}{rgb}{0,0,0.9}
\definecolor{linkblue}{rgb}{.1,.4,.9}
\definecolor{linkgreen}{rgb}{0,0.45,0}
\definecolor{linkpurple}{rgb}{0.7,0.0,0.4}
\definecolor{linkorange}{rgb}{0.7,0.1,0.0}
\definecolor{urlblue}{rgb}{0,0,0.9}
\usepackage[colorlinks=true,
            linkcolor=linkblue,
            citecolor=linkblue,
            urlcolor=linkblue]{hyperref}

\usepackage{url}
\usepackage{orcidlink}

\biboptions{numbers,sort&compress}

\begin{document}
\hypersetup{linkcolor=linkblue, citecolor=linkblue, urlcolor=urlblue}

\begin{frontmatter}

\title{Joint Curvature and Growth Rate Measurements with Supernova Peculiar Velocities and the CMB}

\author[ppgcosmo,fr]{Camilo Crisman\orcidlink{0009-0008-4662-9599}}
\ead{camilo.crisman@edu.ufes.br}

\author[ppgcosmo,cbpf,ov]{Miguel Quartin\orcidlink{0000-0001-5853-6164}}
\ead{mquartin@cbpf.br}

\author[cbpf,ua]{João Rebouças\orcidlink{0000-0002-1667-6019}}
\ead{joaoreboucas@cbpf.br}

\address[ppgcosmo]{PPGCosmo, Universidade Federal do Espírito Santo, 29075-910, Vitória, ES, Brazil}
\address[cbpf]{Centro Brasileiro de Pesquisas Físicas, 22290-180, Rio de Janeiro, RJ, Brazil}
\address[ov]{Observatório do Valongo, Universidade Federal do Rio de Janeiro, 20080-090, Rio de Janeiro, RJ, Brazil}
\address[fr]{Aix Marseille Université, CNRS/IN2P3, CPPM, Marseille, France}
\address[ua]{Department of Astronomy/Steward Observatory, University of Arizona, 933 North Cherry Avenue, Tucson, AZ 85721, USA}

\date{\today }


\begin{abstract}
Type Ia supernova (SN) magnitudes present correlations due to the fact that their peculiar velocities are sourced by the large-scale structure of the Universe. This effect can be used to constrain properties related to the distribution and growth of matter perturbations. We analyze both Pantheon+ and Dark Energy Survey (DES-Y5 and DES-Dovekie) SN catalogues in combination with CMB data from Planck PR4 to constrain $\sigma_8$ in $\Lambda$CDM, optionally including both curvature and a modified growth index $\gamma$.
We show that SN and CMB datasets are highly complementary and capable of measuring $\sigma_8$, $\gamma$ and $\Omega_k$ simultaneously.  Using only SN, we find $\sigma_8 = 0.73 \pm 0.22$ ($0.70_{-0.38}^{+0.31}$) [$1.02^{+0.38}_{-0.45}$] for Pantheon+ (DES-Y5) [DES-Dovekie] in the base flat $\Lambda$CDM model. Interestingly, allowing for free $\gamma$ and $\Omega_k$, we find hints of positive curvature: $\Omega_k = -0.011 \pm 0.006$ $(-0.013^{+0.005}_{-0.006})$ $[-0.008^{+0.004}_{-0.006}]$, which exclude flatness at 2.0$\sigma$ (2.6$\sigma$) [1.8$\sigma$], for the combination of CMB with Pantheon+ (DES-Y5) [DES-Dovekie]. Such hints do not degrade if we also include a modified amplitude of CMB lensing, parametrized by $A_L$. We find that $\gamma = 0.519^{+0.061}_{-0.099}$ ($0.500^{+0.054}_{-0.098}$) [$0.525^{+0.074}_{-0.073}$], which are consistent with the predictions of General Relativity. Finally, the strong degeneracy between all three $\Omega_k$, $\gamma$ and $H_0$ results in a broader CMB $H_0$ posterior. However, if we include SH0ES $H_0$ data, which is in known strong tension with the CMB in flat $\Lambda$CDM, we find that the $H_0$ tension is recast in terms of a significantly negative curvature and suppressed growth of structures.
\end{abstract}


\end{frontmatter}


\section{Introduction}

The vast amount of observational data in current cosmology warrants sophisticated and creative techniques to extract maximal information about theory parameters. In the present day, type Ia supernovae (SN) are still leading observables, able to constrain the background parameters such as the current matter density $\Omega_m$, the spatial curvature parameter $\Omega_k$ and the dark energy equation of state $w$, with percent-level precision~\cite{Pan-STARRS1:2017jku, Brout:2021mpj, Scolnic:2021amr, Rubin:2023jdq, Hoyt:2026fve, DES:2024jxu, DES:2025sig}. In combination with the Cosmic Microwave Background (CMB) power spectra~\cite{Planck:2018vyg, Tristram:2023haj, AtacamaCosmologyTelescope:2025blo, SPT-3G:2025bzu}, as well as distances inferred from the Baryonic Acoustic Oscillation features~\cite{eBOSS:2020yzd, DES:2026aht, DESI:2025zgx}, one can also constrain the amplitude and slope of the primordial power spectrum $P(k)$, the variance of the linear matter field $\sigma_8$, the current baryon density $\Omega_b$, and the optical depth to reionization $\tau$. These multiple probes provide a rather complete understanding of the Universe, its expansion history, and may be used to test possible non-standard physics such as dynamical dark energy and modified gravity~\cite{Planck:2015bue, DESI:2025fii}.

Besides its use as distance indicators, supernovae can be good tracers of Peculiar Velocities (PV) in cosmology. Their PV closely trace those of their host galaxies, perturbing the observed redshift and therefore the inferred luminosity distance $d_L$. Crucially, peculiar velocities are not randomly distributed. Instead, they arise from gravitational infall toward overdense regions and therefore trace the underlying large-scale distribution of matter in the Universe. This connection allows SN peculiar velocity measurements to probe the growth of cosmic structures and to constrain cosmological parameters related to the matter clustering amplitude and growth rate~\cite{Hui:2005nm}.

Peculiar velocities can, especially in combination with other probes, also be used to test new physics scenarios affecting both the spatial geometry and the growth of matter perturbations, such as curvature and modified gravity~\cite{Kim:2019kls}. In this context, an important quantity is the growth rate of structure, defined as $f(a) \equiv d \ln D / d \ln a$, where $D(a)$ is the linear growth factor and $a$ is the scale factor. It is frequently approximated by $f(a) \simeq \Omega_m(a)^{\gamma}$, where $\gamma$ is called the growth index, often assumed to be a constant free parameter~\cite{Amendola:2004wa, Linder:2005in}. This has been shown to be a reasonable approximation for many models with alternative gravitational physics. In flat $\Lambda$CDM this growth rate can be accurately approximated by $\gamma = 0.55$. This is often stated as a general prediction of General Relativity (GR), since the effects of curvature or different equations of state are often small~\cite{Gong:2009sp}. Alternatively, instead of using the $\gamma$ parameter, one can directly measure the product $f(z) \sigma_8(z)$ in different redshift bins, as originally proposed in~\cite{Song:2008qt}.

After the early paper~\cite{Gordon:2007zw}, the interest in SN peculiar velocities has steadily increased, especially in the last decade. The idea of measuring the SN PV correlations has been tested first in~\cite{Castro:2015rrx} and later in~\cite{Macaulay:2016uwy}, which were able to obtain constraints on $\sigma_8$ and, in the former case, also in $\gamma$. Measurements of $f \sigma_8$ with SN were also obtained by~\cite{Huterer:2016uyq, Boruah:2019icj, LSSTDarkEnergyScience:2025irx}. It has been shown that next-generation SN surveys, in particular the Zwicky Transient Facility (ZTF)~\cite{Masci:2018neq,Dhawan:2021hbt}, the Hawai‘i Supernova Flows~\cite{Do:2024iuw} and the Rubin Observatory Legacy Survey of Space and Time (LSST)~\citep{LSSTScience:2009jmu}, will be individually able to measure the SN PV to good precision~\cite{Howlett:2017asw, Graziani:2020kkr}. The correlations of SN PV and DESI galaxy density surveys were more recently measured in~\cite{Nguyen:2025gfc}. Although currently limited to low-redshifts, it was also shown that SN PV will be measurable in next-generation SN surveys to redshifts of at least $z \sim 0.4$~\cite{Garcia:2020qah, Quartin:2021dmr}, and that they are particularly promising when combined with traditional density clustering methods in the same redshift ranges~\cite{Quartin:2021dmr,Amendola:2019lvy,Stahl:2021mat}.
More recently, more attention has been given to SN PV in building the SN catalogs~\cite{Peterson:2021hel,Carr:2021lcj,Carreres:2024rji}.

Being able to probe $\Omega_k$, $\sigma_8$ and $\gamma$ independently from the CMB, supernovae provide a powerful way to test both the amplitude of matter clustering and the laws governing the growth of cosmic structures. This is of particular importance given the current tensions in the field (see~\cite{CosmoVerseNetwork:2025alb} for a review). In particular, as originally shown in~\cite{Gong:2009sp} there is a phenomenological interplay between $\gamma$ and $\Omega_k$, which corresponds to large degeneracies in CMB data. As another example, it was recently shown that the degeneracy between $\gamma$ and the sum of neutrino masses hints that the preference for small or even negative neutrino masses in the DESI data~\cite{Elbers:2025vlz} may indicate either a different value of $\gamma$~\cite{Giare:2025ath} or a non-zero curvature~\cite{Chen:2025mlf}.

In this work, we assess constraints on cosmological parameters obtained using SN PV from the Pantheon+~\cite{Scolnic:2021amr} and the Dark Energy Survey 5 Year catalog (DES-Y5)~\cite{DES:2024jxu, DES:2025sig} catalogues, either by themselves or in combination with Planck CMB data. The DES-Dovekie~\cite{DES:2025sig} catalogue is used in selected representative analyses to provide additional points of comparison between different supernova samples. We assume two theoretical scenarios: $\Lambda$CDM and an extension including both curvature, parametrized by $\Omega_k$, and modifications to the growth of structure, parametrized by the growth index $\gamma$. Incidentally, due to their complementary constraining power, the peculiar velocity field is also being currently probed by different groups using large-scale structure datasets such as Cosmicflows~\cite{Tully:2022rbj}, SDSS~\cite{Qin:2024gra} and DESI~\cite{Said:2024pwm}, where measurements of $f \sigma_8$ have been reported~\cite{Qin:2025rwz, Turner:2025xpy, Bautista:2025ult, Lai:2025xkf}. Here, however, we focus on the current constraints coming from the combination of CMB and SN data, and leave a joint analysis including galaxy surveys for future work.

\section{Methodology}
\label{sec:methodology}
\subsection{SN Peculiar Velocities}

We make use of two compilations of Type Ia supernovae:  Pantheon+~\cite{Scolnic:2021amr} and DES-Y5~\cite{DES:2024jxu}. Pantheon+ consists of approximately 1550 spectroscopically confirmed SN spanning the redshift range $0.001 < z < 2.3$, combining data from multiple surveys and providing one of the most important datasets for cosmological analyses to date. The DES-Y5 sample, obtained from the Dark Energy Survey over five years of observations, includes of order 1800 SN in the range $0.025 < z < 1.13$, with a focus on homogeneous photometry and well-controlled systematics. At low redshift, DES-Y5 and Pantheon+ share a subset of supernovae, leading to a partial overlap between the two samples. In addition, we also analyse the changes in the newer DES-Dovekie catalog, an updated catalog aimed at achieving a better calibration of the DES-Y5 data~\cite{DES:2025sig}.

Following~\cite{Castro:2015rrx}, we model the Type Ia supernova data using a Gaussian likelihood that accounts for correlated uncertainties induced by peculiar velocities. The likelihood is given by
\begin{equation}
    \mathcal{L_{\rm PV}}\propto\vert C \vert^{-1/2}\exp\left[-\frac{1}{2}\delta_m^T C^{-1}\delta_m \right]\,,
\end{equation}
where $\delta_m \equiv \mu(\boldsymbol{\lambda_c}) - \mu^{\rm obs}$ denotes the vector of residuals in distance modulus, which depends on the set of cosmological parameters $\boldsymbol{\lambda_c}$, and $C$ is the total covariance matrix.

The Pantheon+ and DES-Y5 supernova catalogs rely on the empirical light-curve models SALT2~\cite{Brout:2021mpj} and SALT3~\cite{Kenworthy:2021azy, Taylor:2023bag}, respectively, to fit the observed SN properties, including color, peak brightness, and light-curve shape. In both cases, the catalogs implement the BEAMS with Bias Corrections~\cite{Kessler:2016uwi} framework to determine the nuisance parameters associated with color and stretch, effectively calibrating the light-curve standardization. The resulting distance moduli provided by the catalogs are bias-corrected and cosmology-independent. For this reason, in our analysis we directly use the published magnitudes and do not fit the light-curve parameters \(\alpha\), \(\beta\) and \(\gamma\).

The information of the linear peculiar velocity field is encoded in the covariance matrix $C$, which we decompose as
\begin{equation}
    C(\boldsymbol{\lambda_c}, \sigma_v)\,=\,C^\mathrm{PV}(\boldsymbol{\lambda_c})+C^\mathrm{nonlin}({\sigma_v})+C^\mathrm{cat}\,,
\end{equation}
where $C^\mathrm{PV}$ encodes correlations due to peculiar velocities in linear perturbation, $C^\mathrm{nonlin}$ accounts for the non-linear peculiar velocity contributions, and $C^\mathrm{cat}$ accounts for other contributions such as measurement errors, intrinsic magnitude scatter, and calibration uncertainties. The cosmological parameters $\boldsymbol{\lambda_c}$ and the non-linear velocity nuisance parameter $\sigma_v$ are described below.

The contribution from peculiar velocities is derived within linear perturbation theory. The velocity–velocity correlation function between two supernovae located at comoving positions $x_i$ and $x_j$ and observed at redshifts $z_i$ and $z_j$ is defined as
\begin{equation}
\begin{aligned}
    \xi _{ij}^{\text{vel}} &\equiv \langle (\mathbf{v}_{i} \cdot \hat{\mathbf{x}}_{i}) (\mathbf{v}_{j} \cdot \hat{\mathbf{x}}_{j})\rangle \\
    & \!\!\!\!\!\!\! = \!\int\! \frac{d^{3} k}{( 2\pi )^{3}}\frac{(\mathbf{k} \cdot \mathbf{x}_{i})(\mathbf{k} \cdot \mathbf{x}_{j})}{k^{4}} D'_{i} D'_{j} P( k) e^{-i\mathbf{k}(\mathbf{x}_{i} -\mathbf{x}_{j})},
\end{aligned}
\end{equation}
where $D(a)$ is the growth factor of matter perturbations, primes denote derivatives with respect to conformal time $\eta$, and $P(k)$ is the linear matter power spectrum evaluated today. The peculiar velocity covariance matrix elements are then given by
\begin{equation}
\begin{aligned}
    C^\mathrm{PV}_{ij} &=\left(\frac{5}{\ln 10}\right)^{2}\left[ 1-\frac{( 1+z_{i})^{2}}{H( z_{i}) d_{L}( z_{i})} \mathcal{C}_{k}( \chi _{i})\right] \\
    &\! \left[ 1-\frac{( 1+z_{j})^{2}}{H( z_{j}) d_{L}( z_{j})} \mathcal{C}_{k}( \chi _{j})\right] \xi _{ij}^{\text{vel}}( z_{i} ,\mathbf{x}_{i} ,z_{j} ,\mathbf{x}_{j}),
\end{aligned}
\label{eq:covmat}
\end{equation}
where $H(z)$ is the Hubble parameter, $d_L$ is the luminosity distance, and $\chi\equiv\int _{0}^{z}dz'/H( z')$ denotes the comoving radial distance. The function $\mathcal{C}_{k}$ depends on the spatial curvature, $\mathcal{C}_{k} = \cosh \left[H_{0}\sqrt{\Omega _{k}} \chi\right]$.

We compute the linear matter power spectrum $P(k)$ using a modified version of CAMB~\cite{Lewis:1999bs}, called CAMB GammaPrime\footnote{\url{https://github.com/MinhMPA/CAMB_GammaPrime_Growth}}~\cite{Nguyen:2023fip}.
The growth factor and its derivative are consistently computed for each cosmological model sampled in the inference.

As indicated above, the peculiar velocity covariance is separated into two contributions: a linear $C^{\rm PV}$ and a non-linear component $C^{\rm nonlin}$. The latter is modeled through a simple parametrization, represented by a diagonal matrix whose elements are given by
\begin{equation}
    C_{ii}^{\rm nonlin}(\sigma_v)\,=\,\left[\frac{5\sigma_v}{z\ln10} \right]^2\,,
\end{equation}
where we will leave $\sigma_v$ as a free nuisance parameter to be fit together with the cosmological parameters. This covariance accounts for a random component associated with non-linear velocity dispersions including, but not limited to, the rotation velocities of the SN around the host-galaxy.
The covariance matrices provided with the Pantheon+ and DES-Dovekie catalogs include a contribution accounting for non-linear peculiar velocity dispersion, modeled as an effective velocity uncertainty of order $240\mathrm{km/s}$. In contrast, the DES-Y5 covariance matrix does not explicitly include a fixed fiducial velocity-dispersion term. Since our analysis consistently models peculiar velocities, we ensure a uniform treatment across samples by incorporating this effect directly in our likelihood. To avoid any potential double counting in Pantheon+ and DES-Dovekie, we remove the fiducial velocity-dispersion contribution assumed in its covariance construction and replace it with a coherent peculiar-velocity covariance term computed using the value of $\sigma_v$ sampled by our pipeline.

In our baseline analysis $\sigma_v$ is assumed to be constant with redshift. To test this assumption, we considered an extended model for the Pantheon+ sample in which $\sigma_v$ is allowed to vary between two redshift bins, introducing independent parameters $\sigma_{v1}$ and $\sigma_{v2}$ for the ranges $z<0.05$ and $0.05<z<0.1$, respectively. We find that this additional freedom does not lead to significant shifts in the inferred cosmological parameters. Furthermore, despite the extra parameterization, the fit is not sufficiently improved to justify the increased model complexity, and the single $\sigma_v$ model is strongly preferred according to the Bayesian Information Criterion.

We stress that in the likelihood evaluation we vary all parameters simultaneously both in $\mu$ and in the covariance $C$. At each step we therefore recompute a new vector of residuals.

\begin{figure}[t]
    \centering
    \includegraphics[width=1\linewidth]{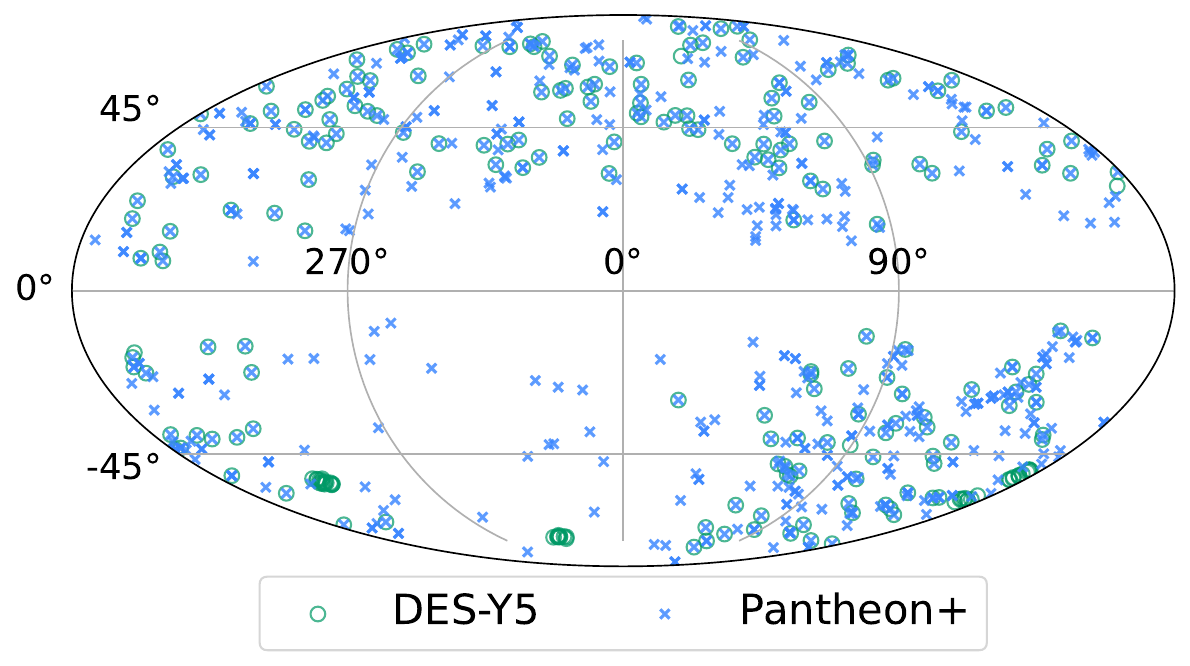}
    \caption{Spatial distribution, in galactic coordinates, of the low-redshift SN samples of Pantheon+ (blue) and DES-Y5 (green) used in our peculiar velocity analysis.}
    \label{fig:mollweide}
\end{figure}

We use Type Ia supernovae from the Pantheon+ and DES-Y5 compilations. For the peculiar velocity analysis we restrict the sample to redshifts $z<z_{\rm max}^{\rm PV}$, where $z_{\rm max}^{\rm PV}$ is 0.1 for Pantheon+ and $0.2$ for DES-Y5 and DES-Dovekie, corresponding to 628, 243 and 245 SNe, respectively. Although future SN catalogs are expected to have relevant peculiar velocity information to higher redshifts~\cite{Garcia:2020qah,Quartin:2021dmr}, this requires a much larger number of supernovae than currently available. We have explicitly tested that for the current catalogues, the information from the velocity-induced correlations saturate below this threshold. Figure~\ref{fig:mollweide} illustrates the spatial distribution of the $z<z^{\text{PV}}_{\text{max}}$ SN from the Pantheon+ and DES-Y5. We note that a fraction of the events is shared between both catalogues, and that DES-Y5 contains many clumped patches in the southern galactic latitudes, corresponding to the foci of the DES SN survey.

\begin{figure}[t!]
    \centering
    \includegraphics[width=.75\linewidth]{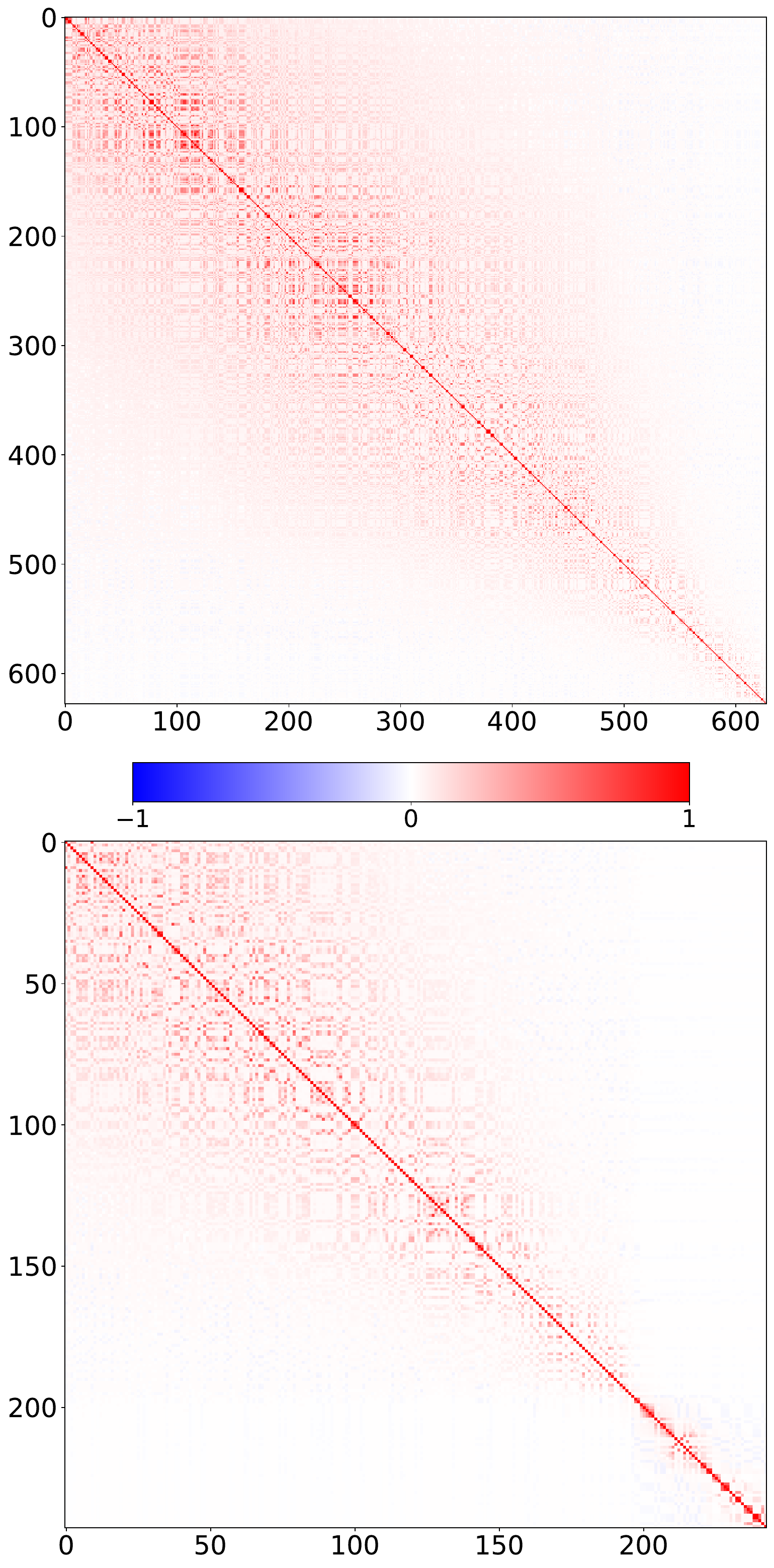}
    \caption{Correlation matrices corresponding to the covariance $C^{\text{PV}}$ of Pantheon+ (top panel) and DES (bottom panel) low-redshift ($z < z_{\rm max}^{\rm PV}$) SN. The SN are ordered from the lowest $z$ up to $z_{\rm max}^{\rm PV}$.
    }
    \label{fig:corr_matrices}
\end{figure}

In contrast, the inference of the background expansion is performed using the full supernova dataset. In order to make this combination, we simply set by hand $C^{\rm PV} = 0$ for all entries where at least one SN has $z > z_{\rm max}^{\rm PV}$. The distance moduli, redshifts, sky positions, and observational covariance matrices are taken directly from the public releases of the respective catalogues. The catalogue covariance matrix $C^{\rm cat}$ includes contributions from photometric uncertainties, intrinsic dispersion, calibration systematics, and selection effects, following the prescriptions provided by the Pantheon+ and DES collaborations. Figure \ref{fig:corr_matrices} displays the correlation matrices derived from the covariance matrices $C^{\text{PV}}$ for the Pantheon+ (top) and DES-Y5 (bottom) $z<z^{\text{PV}}_{\text{max}}$ subsamples. As can be seen, correlations are mostly positive, and extend far outside the main diagonal.

\subsection{CMB Data}

To complement the supernovae peculiar velocity data, we use CMB temperature and polarization anisotropies from the Planck collaboration, specifically the PR4 release implemented in the \texttt{HiLLiPoP} and \texttt{LoLLiPoP} likelihood codes~\cite{Tristram:2023haj}. The \texttt{HiLLiPoP} likelihood accounts for TT, TE and EE power spectra in the multipole range $\ell \in [30, 2500]$ for TT and $\ell \in [30, 2000]$ for TE and EE, while the \texttt{LoLLiPoP} likelihood accounts for the EE power spectrum in the multipole range $\ell \in [2, 30]$. These are supplemented by the \texttt{Commander} likelihood, which provides the TT power spectrum for $\ell \in [2, 30]$~\cite{Planck:2018vyg}. In addition to the primary CMB anisotropies, we also include CMB lensing data from Planck PR3~\cite{planck_pr3_lensing}. In the following, this dataset will be referred to simply as ``CMB''.

The previous iteration of the Planck dataset, namely PR3, presented a so-called lensing anomaly: the data would prefer an amplitude of the lensing effect higher than the $\Lambda$CDM predicted amplitude~\cite{Planck:2018vyg}. This anomaly was modelled by a phenomenological parameter $A_L$, which controls the strength of the lensing effect on the CMB, with $A_L = 1$ denoting the standard effect~\cite{Calabrese:2008rt}. While the Planck PR3 dataset would prefer $A_L = 1.180 \pm 0.065$, the newer PR4 dataset implemented in the \texttt{HiLLiPoP} and \texttt{LoLLiPoP} likelihoods are consistent with standard lensing, with $A_L = 1.039 \pm 0.052$. In the following analysis, for completeness we also consider scenarios where $A_L$ is left as a free parameter.

\subsection{Data Analysis}
We use Markov Chain Monte Carlo (MCMC) techniques to sample the posterior distribution. We use \texttt{Cobaya}~\cite{Torrado:2020dgo} as an interface to the CMB likelihood. To sample the posterior distribution of the cosmological parameters we use both the affine-invariant ensemble sampler \texttt{emcee}~\cite{Foreman-Mackey:2012any}\footnote{\url{https://emcee.readthedocs.io/en/stable}} and the simple Metropolis-Hastings algorithm implemented in \texttt{Cobaya}. We used the former sampler for the SN posterior, and the latter for both the CMB and the combined CMB+SN cases. In \texttt{emcee}, we consider the chain converged when it runs for at least 50 autocorrelation times; in Metropolis-Hastings, we use the Gelman-Rubin criterion with $|R-1| < 0.03$.

We sample over the standard cosmological parameters. Additionally, we consider three extended models parametrized by: $\Omega_k$, which describes the curvature of the Universe, $\gamma$, which accounts for modifications in the growth of matter perturbations, and $A_L$, which controls the amplitude of the lensing effect on the CMB. In summary, we use $\boldsymbol{\lambda_c} \equiv \{h, \Omega_m, \Omega_b, \Omega_k, \sigma_8, n_s, \tau, \gamma\}$ with flat, uninformative priors.  We also sample over nuisance parameters $\boldsymbol{\lambda_n}$ describing supernova (a total of two) and CMB systematics (a total of 18 parameters). Note that when analysing CMB data we do not sample directly over $\Omega_m$, $\Omega_b$, $H_0$ and $\sigma_8$. Instead, as is usually done, we sample over $\omega_c \equiv \Omega_c h^2$, $\omega_b \equiv \Omega_b h^2$, $\theta_{\rm MC}$ (the angular size of the sound horizon at the last scattering surface) and $\log(10^{10}A_s)$, since these are the parameters the CMB data is more sensitive to.  The computational cost of the likelihood evaluations are around 35 (55) seconds for the DES-Y5 (Pantheon+) SN likelihood, and around 8 seconds for the CMB likelihood, which is implemented through Cobaya. In all cases, MCMC convergence took a couple of weeks with parallel processing.

Due to the degeneracy between the supernovae absolute magnitudes in the $B$-band $M_B$ and $H_0$, type Ia supernovae alone are unable to constrain the Hubble parameter. This degeneracy can be broken by adding external data to calibrate the supernovae magnitudes, effectively adding them to the distance ladder. However, the current high-significance of the Hubble tension between the CMB and the local expansion rate measurements (see~\cite{H0DN:2025lyy} for a recent review) means that combining local $H_0$ measurements with CMB data requires a very careful interpretation of the results. Our main analysis therefore does not include $H_0$ measurements, and make use of a broad $H_0$ prior. We nevertheless also analyse the effects of including the $H_0$ prior $\mathcal{N}(73.30, 1.04)$ based on SH0ES~\cite{Riess:2021jrx} in order to test how much the tension is alleviated when we add extra parameters such as curvature, $\gamma$ and $A_L$.

When using SN without CMB, we also include a Gaussian prior on $\omega_b \equiv \Omega_b h^2$ corresponding to the current Big Bang Nucleosynthesis (BBN) constraints obtained in~\cite{Schoneberg:2024ifp}.

The full set of priors used in our analysis is summarized in Table~\ref{tab:priors}.

\begin{table}[t]
\centering
\caption{Prior distributions for model parameters. $\mathcal{U}(a,b)$ denotes a uniform, top-hat, prior from $a$ to $b$, and $\mathcal{N}(\mu,\sigma)$ denotes a Gaussian with standard deviation $\sigma$.
}
\begin{tabular}{l l}
\toprule
Parameter & Prior 
\\
\midrule
$\Omega_k$       & $\mathcal{U}(-0.5, 0.5)$ \\
$\omega_c$       & $\mathcal{U}(0.001, 2)$\\
$\omega_b$       & $\mathcal{N}(0.02196, 0.00063) $\\
\multirow{2}{*}{$H_0$ [km/s/Mpc]} & $\mathcal{U}(40,100)$ (default) \\
                 & $\mathcal{N}(73.30, 1.04)$ (if using SH0ES)\\
$n_s$            & $\mathcal{U}(0.91, 1.05)$ \\
$\tau$            & $\mathcal{U}(0.02, 0.09)$ \\
$\sigma_8$       & $\mathcal{U}(0, 2)$ \\
$\gamma$         & $\mathcal{U}(-1,3)$ \\
$A_L$            & $\mathcal{U}(0.5,2)$ \\
$\sigma_v$ [km/s]& $\mathcal{U}(0,450)$ \\

\bottomrule
\end{tabular}
\label{tab:priors}
\end{table}

\section{Results}
\label{sec:results}

\subsection{Flat \texorpdfstring{$\Lambda$}{Lambda-}CDM}

\begin{figure}[t]
    \centering
    \includegraphics[width=\linewidth]{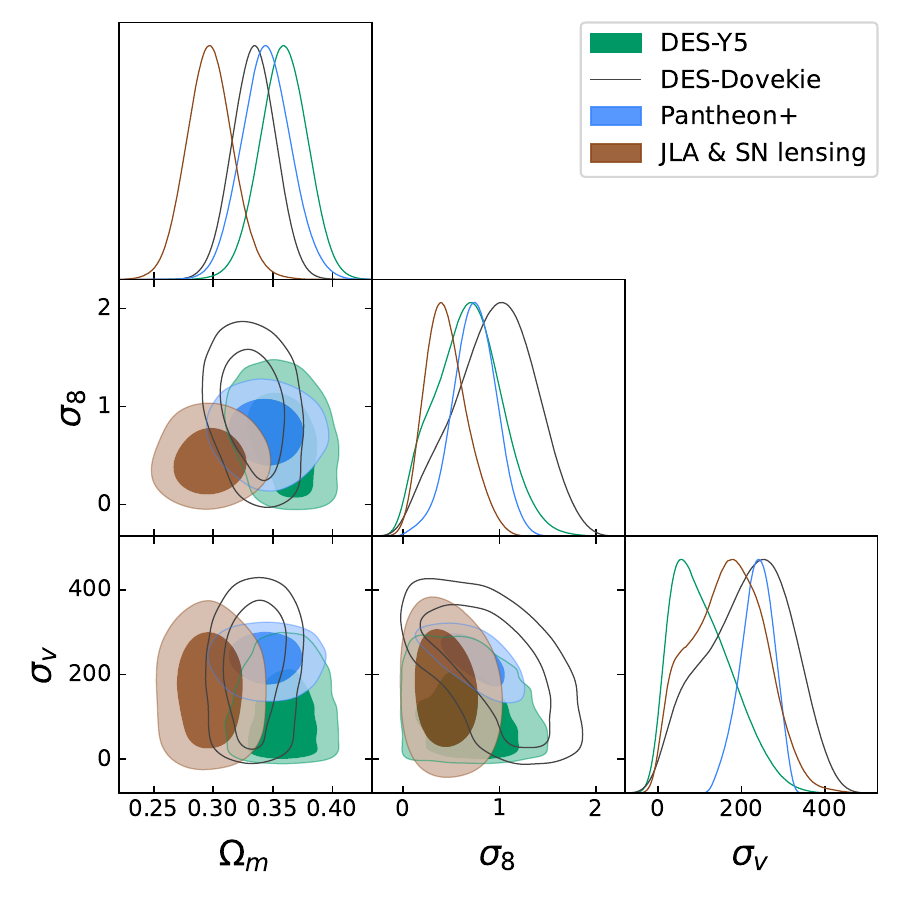}
    \caption{Confidence contours (1 and 2$\sigma$) on the parameters $\Omega_m$, $\sigma_8$ and $\sigma_v$ from supernovae peculiar velocities. Green contours show constraints using DES-Y5 supernovae, whereas blue contours correspond to Pantheon+ and brown contours are taken from \cite{Castro:2015rrx} and represent constraints from JLA combined with SNe lensing.
    }
    \label{fig:sne_sig8}
\end{figure}

We start by analyzing the results obtained exclusively with supernova data, assuming the standard flat $\Lambda$CDM model within GR ($\gamma = 0.55$). Figure~\ref{fig:sne_sig8} depicts the constraints from Pantheon+ and DES-Y5. We also add the constraint obtained in~\cite{Castro:2015rrx} from the older JLA SN catalogue~\cite{SDSS:2014iwm}. In that analysis, SN peculiar velocity and SN lensing data (see~\cite{Quartin:2013moa} for the lensing methodology) were combined to obtain the constraint $\sigma_8 = 0.44 \pm 0.21$. We note that the peculiar-velocity posterior from the newer catalogues have comparable precision to the joint PV and lensing posterior from JLA. This limited precision stems from the small number of new low-redshift supernovae in the newer samples. More importantly, the accuracy of the data appears much improved, yielding $\sigma_8$ measurements which are much more consistent with CMB and galaxy survey measurements: to wit, $\sigma_8 = 0.73 \pm 0.22$ for Pantheon+, $\sigma_8 = 0.70_{-0.38}^{+0.31}$  for DES-Y5 and $\sigma_8 = 1.02^{+0.38}_{-0.45}$ for DES-Dovekie, compared to $\sigma_8 = 0.751^{+0.034}_{-0.036}$ from the DES-Y6 3x2pt analysis~\cite{DES:2026fyc} and $\sigma_8 = 0.8070 \pm 0.0065$ from the Planck PR4 analysis~\cite{Tristram:2023haj}. The low amount of peculiar velocity correlations in the JLA data was also noted in~\cite{Huterer:2015gpa}. These results corroborate the improved PV handling in the Pantheon+ catalogue, discussed in detail in~\cite{Carr:2021lcj}.

The background constraints of these datasets have been compared in the past, in particular their discrepancies in the preferred $\Omega_m$, but their peculiar velocity constraints have not been compared before. Overall, we find a reasonable agreement between Pantheon+, DES-Y5 and DES-Dovekie datasets, but there are some important differences. First, while we do not notice any significant relative bias in the $\sigma_8$ constraints between Pantheon+ and DES-Y5, we find that DES-Dovekie prefers larger $\sigma_8$ values. It is also less precise than DES-Y5. Regarding the amount of non-linear PV dispersions, whereas Pantheon+ data exhibits a Gaussian posterior, allowing only the narrow range $180-320\,$km/s, DES-Y5 prefers lower values, while DES-Dovekie has a broad posterior allowing values in the whole range $0-450\,$km/s.
We note that the BBN priors used here do not significantly affect the $\sigma_8$ constraints.

While it was expected that the DES datasets, due to the lower number of low-redshift supernovae, would exhibit a somewhat lower precision than Pantheon+, the differences between both DES datasets are more surprising. Given the good agreement between Pantheon+ and DES-Y5, we will use both catalogs as our baseline in the rest of the analyses in this work, and leave more detailed comparisons of both DES datasets for future work.

\begin{figure}[t]
    \centering
    \includegraphics[width=1\linewidth]{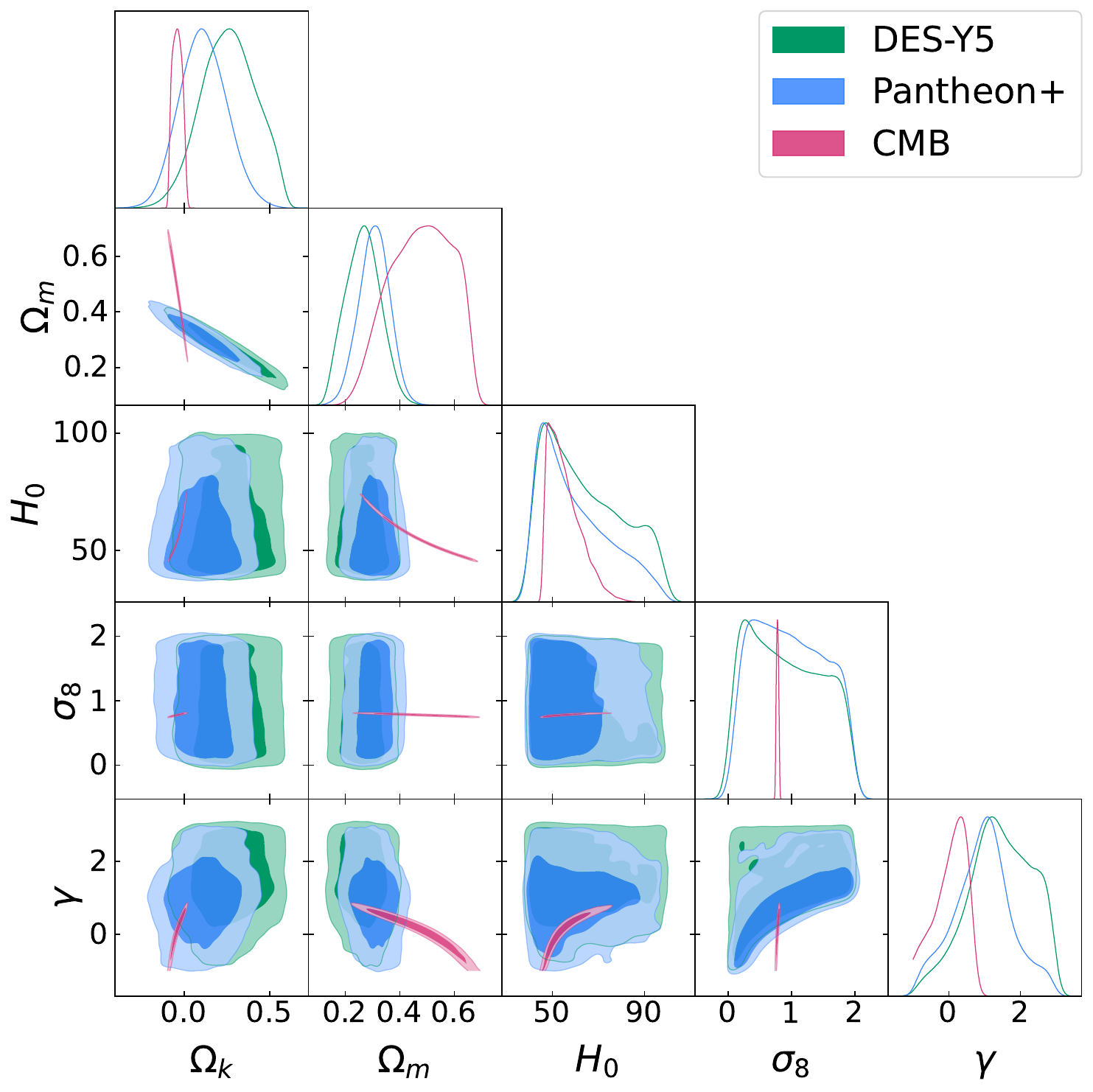}
    \caption{Similar to Figure~\ref{fig:sne_sig8}, but adding CMB data and the  $\Omega_k$ and $\gamma$ parameters. Green and blue contours show constraints using DES-Y5 and Pantheon+, respectively. Purple contours show the CMB results.
    }
    \label{fig:MCMC4}
\end{figure}

\subsection{Curvature and Modified Gravity}

We now turn our attention to the more general model, which includes both a free spatial curvature and a free growth index $\gamma$. In Figure \ref{fig:MCMC4} we present the main cosmological constraints where the blue and green contours correspond to the Pantheon+ and DES-Y5 supernova datasets, respectively, while the pink contours show the constraints obtained from CMB.
As expected, supernova data without an $M_B$ or $H_0$ prior have limited constraining power on $H_0$. Likewise, as anticipated, we note a strong degeneracy between $\sigma_8$ and $\gamma$. On the other hand, when spatial curvature is allowed to vary, the CMB constraints also exhibit multiple degeneracies. In particular, the $\sigma_8 - \gamma$ degeneracy from CMB appears approximately orthogonal to that of the supernova data, highlighting the complementarity between the two probes, first observed in~\cite{Quartin:2021dmr}. This illustrates why a joint analysis of both probes can be powerful.

\begin{figure}[t]
    \centering
    \includegraphics[width=1\linewidth]{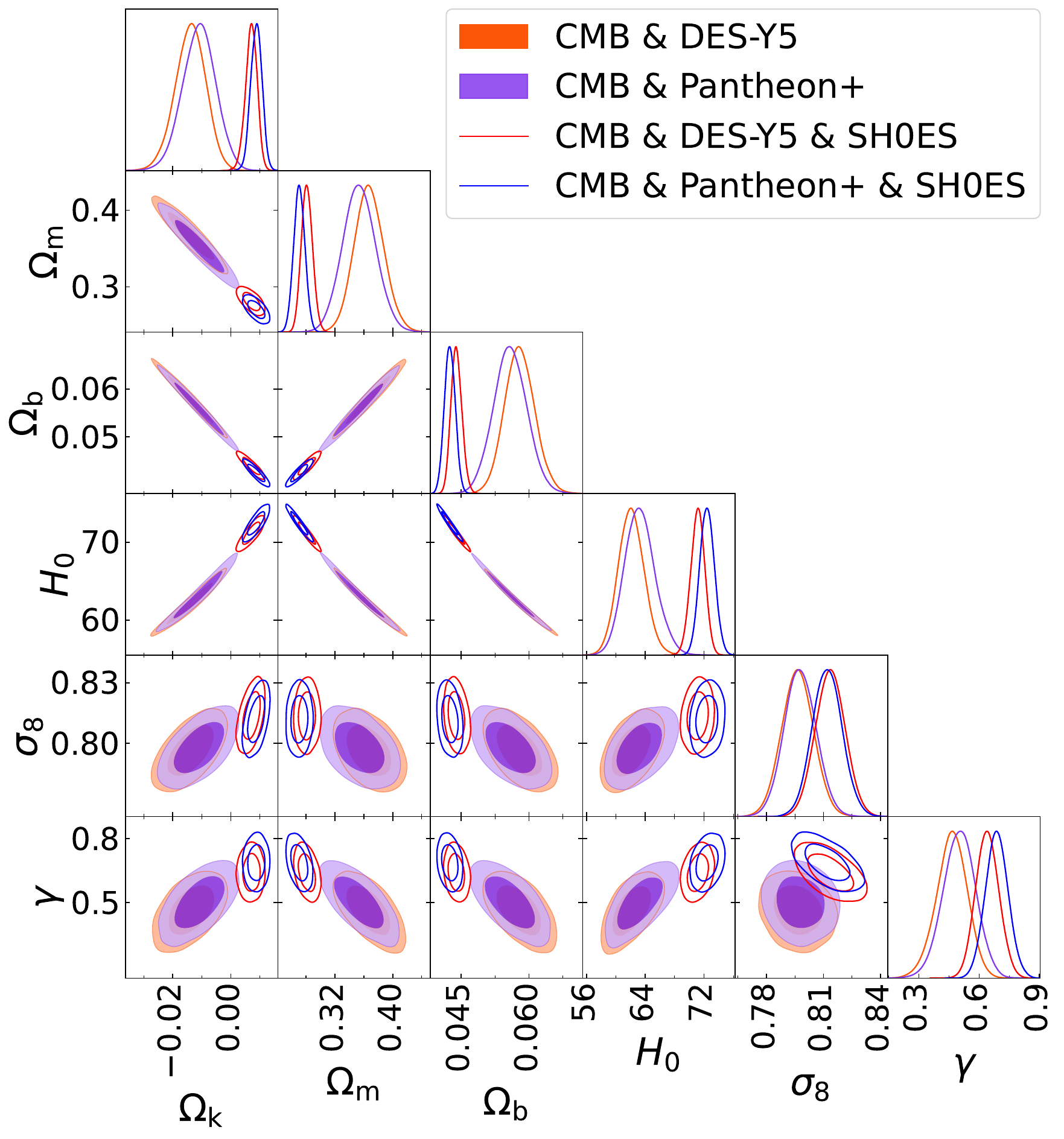}
    \caption{Same as Figure~\ref{fig:MCMC4}, but for the combined CMB and SN posterior. Orange and purple contours show the constraints using CMB \& DES-Y5 and CMB \& Pantheon+, respectively. Filled contours show constraints without the SH0ES $H_0$ prior; unfilled contours use the SH0ES prior (this combination is unreliable, as discussed in the text). The tension between both cases is clear even with the inclusion of a free $\Omega_k$ and $\gamma$.}
    \label{fig:shoeswandwo}
\end{figure}

In Figure~\ref{fig:shoeswandwo} we present the constraints obtained with this joint analysis of SN velocities and the CMB. We first note that, once again, we find a good agreement between the Pantheon+ and DES-Y5 catalogs. Moreover, as anticipated, we find a significant increase in precision in this case. In particular, we get meaningful simultaneous measurements of $\Omega_k$, $\sigma_8$ and $\gamma$. We find a $2.\sigma$ ($2.6\sigma$) [$1.8\sigma$] preference for negative $\Omega_k$ for Pantheon+ (DES-Y5) [DES-Dovekie]:
\begin{itemize}[itemsep=2pt, topsep=5pt]
    \item $\Omega_k = -0.011 \pm 0.006\;$  [CMB \& Pantheon+]\,,
    \item $\Omega_k = -0.013^{+0.005}_{-0.006}\;$  [CMB \& DES-Y5]\,,
    \item $\Omega_k = -0.008^{+0.004}_{-0.006} \;$  [CMB \& DES-Dovekie]\,.
\end{itemize}
The CMB-only results are instead $\Omega_k = -0.042^{+0.031}_{-0.029}$, which can be compared to the official PR4 ones with fixed $\gamma$, to wit $\Omega_k = -0.0078 \pm 0.0058$ (or $\Omega_k = -0.012 \pm 0.010$ without lensing)~\cite{Tristram:2023haj}.
This five-fold decrease in precision is because $\Omega_k$ is highly correlated with $\gamma$, as can be seen in Figure~\ref{fig:MCMC4}, and as was previously illustrated in~\cite{Nguyen:2023fip} (see also \cite{Gong:2009sp} for early discussions on the interplay between both parameters).  When comparing the combined and CMB-only results, we see that the addition of SN velocities results in very similar precision for the free $\gamma$ case to the CMB-only results with a fixed $\gamma$, which is remarkable. Nevertheless, there is a shift of the posterior towards negative $\Omega_k$ by around 1$\sigma$. The same happens for the constraints on $H_0$. We get $H_0 = 63.2 \pm 2.0 \; (61.9^{+2.1}_{-1.5}) \; [63.9^{+2.1}_{-1.5}] \, \mathrm{km/s/Mpc}$ combining CMB with Pantheon+ (DES-Y5) [DES-Dovekie], compared to $63.6^{+2.1}_{-2.3}\,\mathrm{km/s/Mpc}$ from Planck PR3~\cite{Planck:2018vyg} and $64.6\pm2.3\,\mathrm{km/s/Mpc}$ from PR4~\cite{Tristram:2023haj}, both with fixed $\gamma$.

For $\gamma$, we obtain values that are consistent with the GR prediction:

\begin{itemize}[itemsep=2pt, topsep=5pt]
    \item $\gamma = 0.519^{+0.061}_{-0.099}\;$  [CMB \& Pantheon+]\,,
    \item $\gamma = 0.500^{+0.054}_{-0.098}\;$ [CMB \& DES-Y5]\,,
    \item $\gamma = 0.525^{+0.074}_{-0.073}\;$ [CMB \& DES-Dovekie]\,.
\end{itemize}
This is in contrast with what was reported for Planck PR3 data in~\cite{Nguyen:2023fip,Specogna:2023nkq}, but in line with what was found in other works which made use of the \texttt{HiLLiPoP} likelihood and PR4~\cite{Specogna:2024euz}.
This latter work found, assuming flat $\Lambda$CDM and using CMB-only data with \texttt{HiLLiPoP}, the constraint $\gamma = 0.621 \pm 0.090$. We thus see, once more, that by adding SN PV to the CMB data, one can include curvature and still achieve similar precision for $\gamma$ compared to the CMB-only case with one less degree of freedom.

\begin{figure}[t]
    \centering
    \includegraphics[width=1\linewidth]{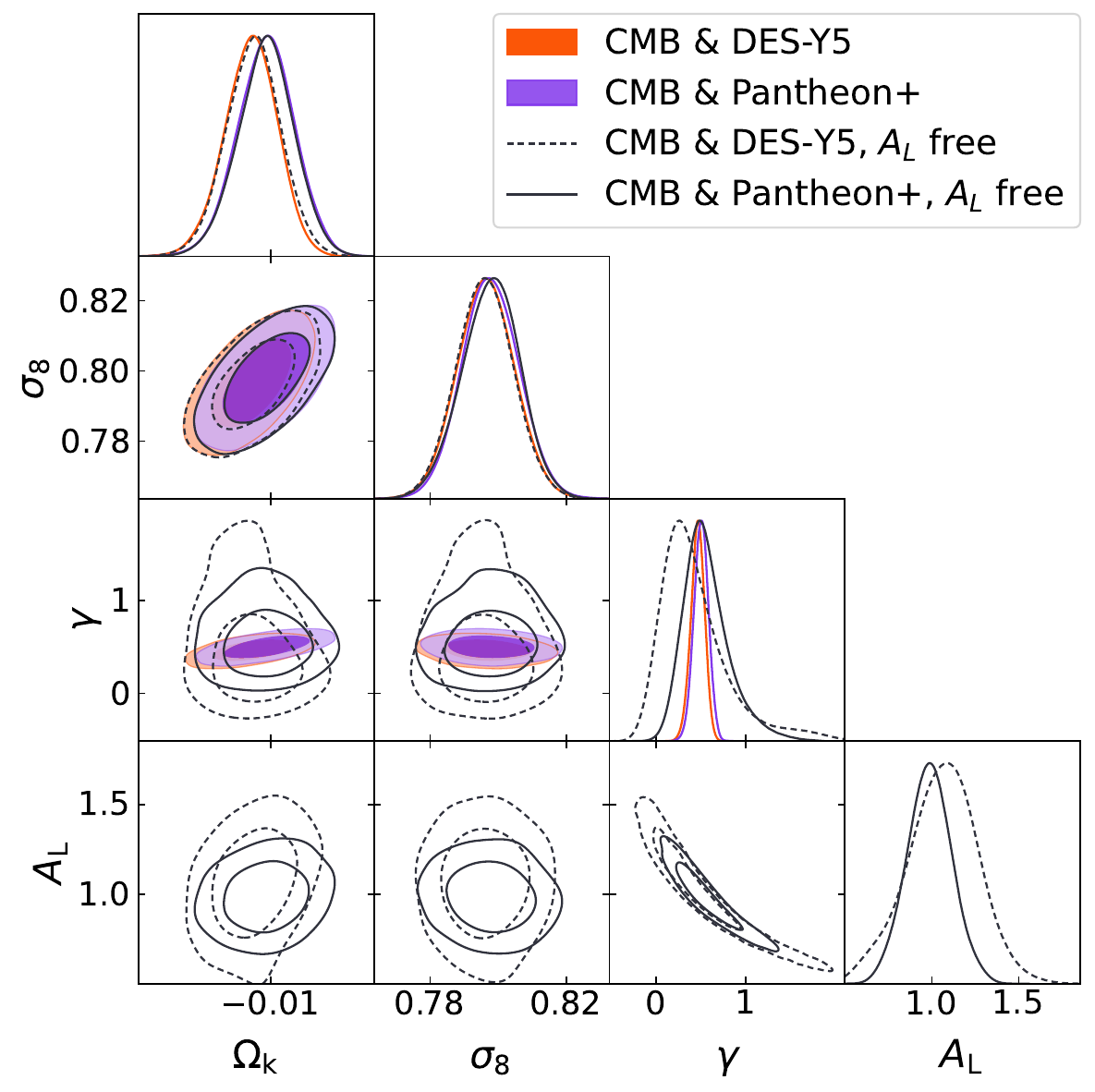}
    \caption{
    Constraints using the joint SN PV + CMB posterior, for both the $A_L \equiv 1$ and free $A_L$ cases. This illustrates the mutual degeneracies between these parameters. We note that the inclusion of a free $A_L$ greatly diminishes the precision of $\gamma$, but does not affect $\sigma_8$ or $\Omega_k$.
    }
    \label{fig:ok_sig8_gamma}
\end{figure}

\begin{figure}[t!]
    \centering
    \includegraphics[trim={0 2.6cm 0 0}, clip, width=.4\textwidth]{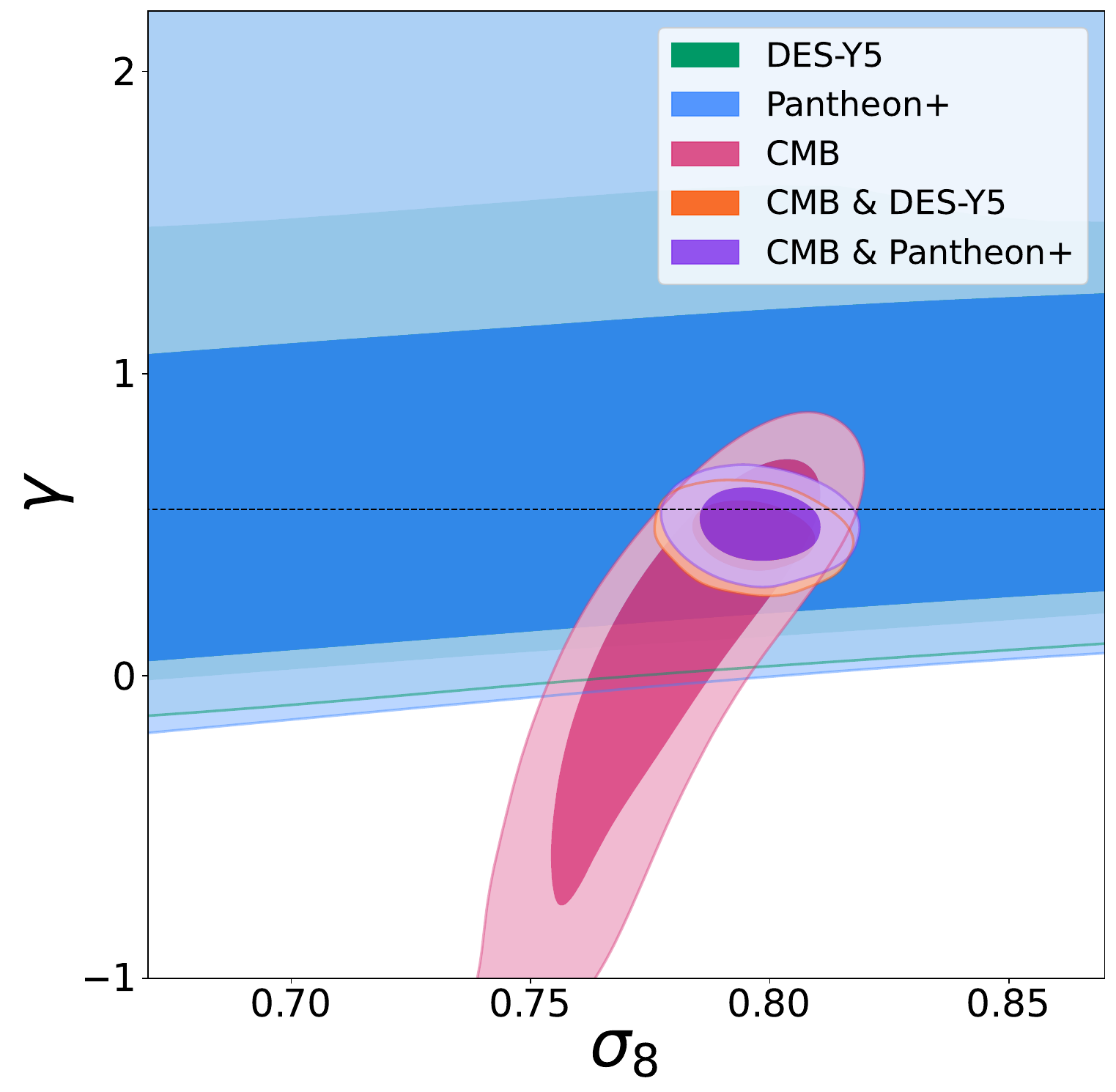}
    \quad \includegraphics[trim={0 0 0 0.27cm}, clip, width=.4\textwidth]{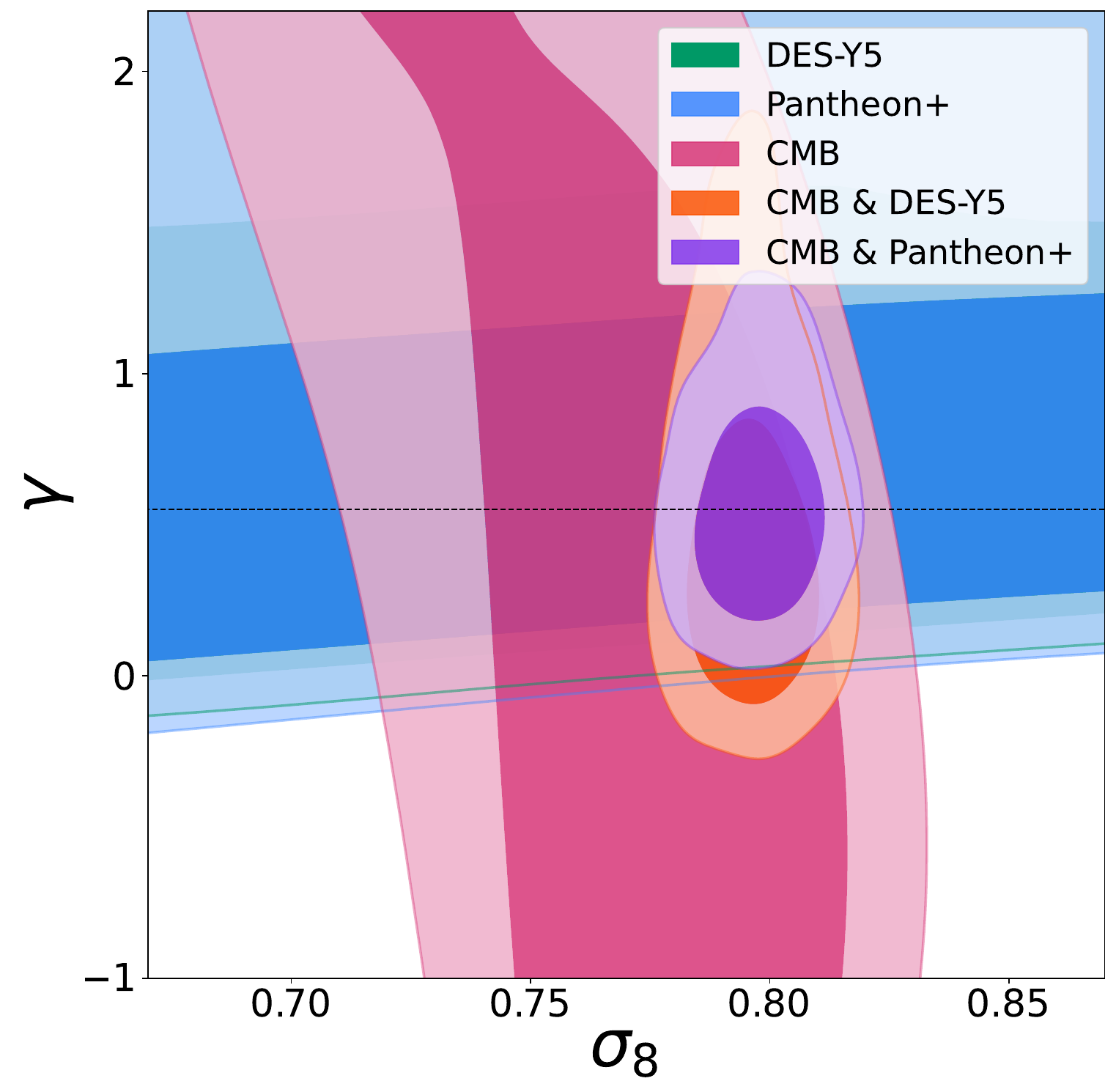}
    \caption{Constraints on $\sigma_8$ and $\gamma$ using CMB, SN peculiar velocities, and their combination. \emph{Top:} $A_L \equiv 1$. \emph{Bottom:} $A_L$ left as a free parameter. In both cases, we observe that peculiar velocities break CMB degeneracies, much improving the final precision.}
    \label{fig:sig8_gamma_combined}
\end{figure}

\begin{table*}[t]
\centering
\caption{Marginalized 1$\sigma$ CMB+PV constraints for the $A_L=1$ case and no SH0ES prior.}
\label{tab:parameter_constraints_transposed}
    \begin{tabular}{lccc}
    \toprule
    Parameter & Pantheon+ & DES-Y5 & DES-Dovekie \\
    \midrule
    $\Omega_k$ & $-0.011 \pm 0.006$ & $-0.013^{+0.005}_{-0.006}$ &  $-0.008^{+0.004}_{-0.006}$ \\
    $\Omega_m$ & $0.35 \pm 0.02$ & $0.36 \pm 0.02$& $0.34 \pm 0.02 $\\
    $\Omega_b$ & $0.056 \pm 0.004$ & $0.058 \pm 0.003$& $0.054\pm0.003 $\\[2pt]
    $H_0$ [$\mathrm{km}/\mathrm{s}/\mathrm{Mpc}$] & 63.2 $\pm$ 2.0 & $61.8^{+2.1}_{-1.5} $ & $63.9^{+2.1}_{-1.5}$ \\[3pt]
    $\sigma_8$ & 0.798 $\pm$ 0.009 & 0.796 $\pm$ 0.009& $0.798^{+0.010}_{-0.006}$ \\[3pt]
    $\gamma$ & $0.519^{+0.061}_{-0.099}$ & $0.500^{+0.054}_{-0.098}$& $0.525^{+0.074}_{-0.073}$\\
    $\sigma_v$ [$\mathrm{km}/\mathrm{s}$]  & 232$\pm$33  &  $65^{+172}_{-57}$ & $251\pm73$  \\
    \bottomrule
    \end{tabular}
\end{table*}

In past CMB analyses, it was shown that a preference for $A_L > 1$ was correlated with hints for $\Omega_k < 0$~\cite{Planck:2018vyg, AtacamaCosmologyTelescope:2025blo, DiValentino:2019qzk}, and the correlations between the two parameters was also explored in~\cite{2026JCAP...01..009W}. In Figure~\ref{fig:ok_sig8_gamma} we illustrate the effect of allowing $A_L$ as a free parameter, focusing on $\Omega_k$, $\sigma_8$ and $\gamma$, with all other parameters marginalized over. Marginalized constraints are shown in Table~\ref{tab:parameter_constraints_complete}. First, we find that the lensing effect amplitude is consistent with the standard prediction:
\begin{itemize}[itemsep=2pt, topsep=5pt]
    \item $A_L = 0.99 \pm 0.13\;$  [CMB \& Pantheon+]\,,
    \item $A_L = 1.07 \pm 0.20\;$ [CMB \& DES-Y5]\,.
\end{itemize}
The $A_L$ parameter is strongly (negatively) correlated with $\gamma$. This is expected, as an increase of $\gamma$ enhances the late-time growth of structures, which can be compensated by a decrease in the lensing effect through a smaller $A_L$. This correlation greatly diminishes the precision of $\gamma$, after marginalization, by a factor around 5. However, it does not shift the $\gamma$ posterior significantly, and most importantly, we observe no significant shifts in the $\sigma_8$, $\Omega_k$ and $H_0$ parameters, which are largely uncorrelated with $A_L$. This demonstrates the robustness of our results to possible internal inconsistencies regarding the lensing effect in the CMB.

In Figure~\ref{fig:sig8_gamma_combined} we present a more detailed analysis of the joint $\sigma_8$ and $\gamma$ contours. The top panel shows the case in which the CMB lensing amplitude parameter $A_L$ is fixed to its standard value $A_L=1$, as in Figure~\ref{fig:shoeswandwo}; the bottom panel corresponds to the case where $A_L$ is treated as a free parameter.
In both panels, the different degeneracy directions for the SN PV and CMB datasets are clearly visible. These degeneracies are related to the correlation between $\Omega_k$ and the growth index $\gamma$, which propagates into the $\sigma_8-\gamma$ parameter space. In the scenario where $A_L$ is allowed to vary freely, the degeneracy in the CMB-only constraints becomes significantly stronger. In particular, for the CMB-only case the extended parameter space induces a near-complete degeneracy that prevents a meaningful determination of $\gamma$. This again highlights the importance of combining CMB data with low-redshift probes such as peculiar velocities.

Table~\ref{tab:parameter_constraints_transposed} summarizes the marginalized $1\sigma$ constraints in each cosmological parameter for the CMB+PV posterior assuming $A_L=1$ case and without adding the SH0ES prior. The Table with the cases with free $A_L$ or with the addition of the SH0ES priors is shown in~\ref{app:tables}.

\subsubsection{SN peculiar velocities and the \texorpdfstring{$H_0$}{H\_0} tension}

We now turn our attention to the Hubble tension and related discrepancies between the CMB and local $H_0$ measurements. We start by noting that the strong CMB degeneracy between all three $\Omega_k$, $\gamma$ and $H_0$, result in a much broader CMB $H_0$ posterior, alleviating the $H_0$ tension significantly, from the 5.0$\sigma$ level reported in~\cite{Riess:2021jrx}. The inclusion of curvature is known to increase the $H_0$ uncertainty, which decreases in turn the tension. For instance, for PR4 TTTEEE+lensing, it results in $H_0 = 64.6 \pm 2.3\,$km/s/Mpc, which corresponds to a 3.5$\sigma$ tension with SH0ES 2022. Here we extend this analysis by including also a free $\gamma$, which results in
\begin{equation}
    H_0 = 49.8^{+8.5}_{-3.6}\,{\rm km/s/Mpc}\,.
\end{equation}
This shrinks the tension between SH0ES and CMB to only 2.2$\sigma$.
One could thus question what would the posterior be when combining SN PV, CMB and the SH0ES $H_0$ prior. However, the joint SN+CMB posterior has much tighter $H_0$ constraints  -- see Table~\ref{tab:parameter_constraints_transposed}. These values remain in strong tension with SH0ES 2022, between $3.9\sigma$ and $4.5\sigma$ (see~\ref{app:tension} for details). Therefore a combination of SN+CMB and SH0ES can yield non-sensical results. Nevertheless, it is revealing to see the results of naively including SH0ES in the joint CMB-SN analysis with the extra degrees of freedom here considered: $\Omega_k$, $\gamma$ and $A_L$.

In Figure~\ref{fig:shoeswandwo} we also show the contours including SH0ES in the case in which $A_L = 1$. We see that the inclusion of the SH0ES prior produces marked shifts, as expected. Firstly, it induces a sign flip in $\Omega_k$: we get $\Omega_k = 0.009 \pm 0.002$ ($0.007\pm0.002$) for Pantheon+ (DES-Y5), which at face value would mean a strong detection of a negative spatial curvature. The value of $\gamma$ also shifts upwards to $0.69\pm0.06 \; (0.64\pm0.06)$ for Pantheon+ (DES-Y5), and become in small tension with the prediction of General Relativity. In particular, the point $\{\Omega_k = 0, \, \gamma = 0.55 \}$, corresponding to the flat GR case, is excluded at over $4.4 \sigma$ (around $3.1\sigma$) for Pantheon+ (DES-Y5).\footnote{Higher significances than 4.4$\sigma$ would require many more effective MCMC points.} This shows that the $5\sigma$ tension between our $H_0$ fiducial and the Planck data is being recast as a $3-4.5\sigma$ exclusion of the $\{\Omega_k = 0, \, \gamma = 0.55 \}$ model. We also observe that the inclusion of the SH0ES prior leads to significantly lower inferred values of $\Omega_b$ and $\Omega_m$.

These shifts are all very apparent in Figure~\ref{fig:shoeswandwo}, which illustrates that the posteriors with and without SH0ES remain in clear tension even with these extra degrees of freedom. The shifts also reflect the non-trivial degeneracies between $H_0$, spatial curvature, and matter density in the joint parameter space. We therefore see that a naive use of the SH0ES prior in a joint SN+CMB posterior results in a likely spurious tension with the flat case with the expected GR growth of structure.

Finally, since leaving $A_L$ free does not change the constraints on the cosmological parameters apart from increasing the error bars on $\gamma$, we conclude that this tension cannot be alleviated by altering the lensing effect on the CMB.

\section{Conclusions}
\label{sec:conclusions}

Peculiar velocities from Type Ia supernovae is an interesting and powerful complementary probe of large-scale structure, and allows SN to transcend its traditional role of only constraining background parameters. Although current catalogues are limited in size, on-going next generation surveys are detecting substantially more low-redshift supernovae. Peculiar velocity measurements are also highly complementary to other cosmological probes, and their combination breaks important degeneracies in cosmological parameters. This allows for advanced analysis with high precision of models with a few more degrees of freedom than $\Lambda$CDM, such as those of curvature, different growth rates, and/or different amounts of weak lensing.

In this work we have demonstrated how a combination of SN velocities and the CMB can already lead to substantial improvements, even in the absence of data from next-generation low-redshift SN surveys. Although the individual probes cannot constrain well neither $\gamma$ or $\Omega_k$, the combined analysis results in meaningful constraints. First, we get $15\%$ precision on $\gamma$ and values which are consistent with the prediction of General Relativity, and $1\%$ constraints on $\sigma_8$. Interestingly, we also obtain precise constraints on $\Omega_k$ that hint (between 2.2 and 3.0$\sigma$) at a positive spatial curvature of the Universe. Alternatively, we also provide independent constraints on $f(z)\sigma_8(z)$ at the 10\% precision level, which is comparable to state-of-the-art measurements using around 11,000 galaxy peculiar velocities~\cite{Lai:2025xkf}.

We have also assessed the effect of the Hubble tension on this scenario with modified gravity and curvature by including a prior on $H_0$ matching the SH0ES results. Allowing for both free $\Omega_k$ and $\gamma$, the tension between Planck PR4 and SH0ES 2022 shrink from 5.0 to 2.2$\sigma$, with the caveat that this is driven by the increased uncertainty and not by significant shifts of the posterior peaks. When the SH0ES $H_0$ prior is included, the remaining tension in $H_0$ propagates into the curvature sector, leading to a mild preference for positive spatial curvature and higher $\gamma$ at $3.1-4.4\sigma$ significance levels. We find that allowing for an additional degrees of freedom $A_L$ for the CMB lensing amplitude does not resolve this behavior, indicating that the observed shifts are another manifestation of the Hubble tension, and disconnected from simple corrections to the CMB lensing.

We found that the peculiar velocity results of both Pantheon+ and DES-Y5 are compatible between themselves and in better agreement with the CMB results than previous supernova compilations such as the JLA catalog, which lends more robustness to this present analysis.

Now that we have established the potential of joint SN peculiar velocity analysis with other probes, more precision can be achieved by combining with other datasets beyond the CMB. We plan to pursue this avenue in future works.

\begin{table*}[!]
\centering
\renewcommand{\arraystretch}{1.2}
\begin{tabular}{lcccccc}
    \toprule
    \multirow{2}{*}{Parameter} & \multicolumn{2}{c}{P+} & \multicolumn{2}{c}{DES-Y5} & P+ & DES-Y5\\
    \cmidrule(lr){2-3} \cmidrule(lr){4-5} \cmidrule(lr){6-6} \cmidrule(lr){7-7} & $A_L=1$ & Free $A_L$ & $A_L=1$ & Free $A_L$ & \multicolumn{2}{c}{(SH0ES $\& \ A_L=1$)}  \\
    \midrule
    $\Omega_k$ & \multicolumn{2}{c}{$-0.011\pm$0.006}  & $-0.013^{+0.005}_{-0.006}$                   & $-0.014\pm$0.006  & 0.009 $\pm$ 0.002 & 0.007$\pm$0.002 \\
    $\Omega_m$ & \multicolumn{2}{c}{0.35$\pm$0.02}     & \multicolumn{2}{c}{0.36$\pm$0.02}  & 0.271$\pm$0.008 & 0.281$\pm$0.008 \\
    $\Omega_b$ & \multicolumn{2}{c}{0.056$\pm$0.004}   & \multicolumn{2}{c}{0.058$\pm$0.003}& 0.042$\pm$0.001 & 0.044$\pm$0.001 \\
    $\sigma_8$ & \multicolumn{2}{c}{0.798$\pm$0.008}   & \multicolumn{2}{c}{0.796$\pm$0.008}& 0.812$\pm$0.008 & 0.813$\pm$0.008 \\
    $H_0$      & $63.2^{+2.1}_{-2.0}$    & $63.2^{+2.0}_{-1.9}$& $61.8^{+2.1}_{-1.5}$   & $62.2_{-1.8}^{+2.0}$& 72.4$\pm$1.0& 71.2$\pm$0.9 \\
    $\gamma$   & $0.519^{+0.061}_{-0.099}$           & $0.53_{-0.21}^{+0.28}$& $0.500^{+0.054}_{-0.098}$         & $0.36_{-0.26}^{+0.41}$& 0.69$\pm$0.06& 0.64$\pm$0.06 \\
    $A_L$      & [1]                     & 0.99$\pm$0.13     & [1]                   & 1.07$\pm$0.20      & [1] & [1]\\
    \bottomrule
\end{tabular}
\caption{Marginalized 1$\sigma$ CMB+PV constraints for the different posterior cases here considered. Note that the case combining CMB and SH0ES includes many caveats due to the $H_0$ tension [see text].
}
\label{tab:parameter_constraints_complete}
\end{table*}

\section*{Acknowledgements}

We thank Julián Bautista for a thorough revision of an earlier draft and for discussions. This study was financed in part by the Coordenação de Aperfeiçoamento de Pessoal de Nível Superior – Brasil (CAPES) – Finance Code 001. CC and JR acknowledge financial support from CAPES.
MQ is supported by the Brazilian research agencies Fundação Carlos Chagas Filho de Amparo à Pesquisa do Estado do Rio de Janeiro (FAPERJ) project E-26/201.237/2022 and CNPq (Conselho Nacional de Desenvolvimento Científico e Tecnológico).
We acknowledge the use of the computational resources of the joint CHE / Milliways cluster, supported by a FAPERJ grant E-26/210.130/2023.

\appendix

\section{Table of constraints}\label{app:tables}

Here we show Table~\ref{tab:parameter_constraints_complete}, with the final $1\sigma$ marginalized constraints on all parameters, for all models analyzed with either Pantheon+ and DES-Y5 catalogs. As discussed in the main text, combining SH0ES with the joint SN+CMB posterior is problematic due to the large tension between both.

\bigskip

\section{The $H_0$ tension including curvature and $\gamma$}\label{app:tension}

\begin{figure}[t!]
    \centering
    \includegraphics[width=.94\linewidth]{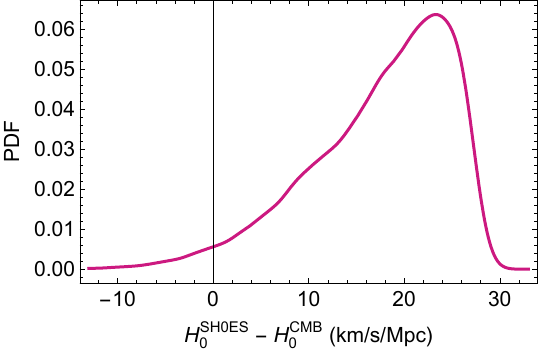}
    \includegraphics[width=.94\linewidth]{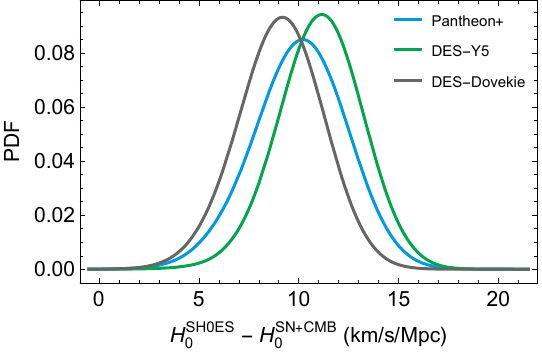}
    \caption{PDF of $H_0^{\rm SH0ES} - H_0^{\rm CMB}$ (top) and $H_0^{\rm SH0ES} - H_0^{\rm SN+CMB}$ (bottom panel) when including both $\Omega_k$ and $\gamma$ degrees of freedom. While for CMB-only the tension is alleviated to only 2.2$\sigma$, for SN+CMB the tension remains large ($\sim 4\sigma$).}
    \label{fig:Delta-PDF}
\end{figure}

Here we illustrate the $H_0$ tension between SH0ES 2022 and either the Planck CMB data alone, or its combination with supernovae, when one includes both $\Omega_k$ and $\gamma$ as free parameters. In all cases, we start with the marginalized MCMC chains on $H_0$ from the joint SN+CMB case. We subsequently draw the same number of random samples from a Gaussian distribution as defined by the SH0ES prior. We then sample a new random variable $\Delta$ defined as $\Delta \equiv H_0^{\rm SH0ES} - H_0^{\rm X}$, where $X$ stands for CMB or CMB+SN, and compute the probability that $\Delta < 0$. Finally, we convert this result into the corresponding Gaussian $\sigma$ level. Figure~\ref{fig:Delta-PDF} illustrates the results, where a smoothing with a narrow kernel is used for visualization purposes.

We see that the inclusion of $\Omega_k$ and $\gamma$ significantly alleviates the tension between CMB data and the local measurements, and the probability to have $\Delta < 0$ is 2.5\%, which corresponds to only a $2.2\sigma$ Gaussian significance. However, as discussed in the main text, combining CMB and SN data breaks important degeneracies, which lead to a significant tightening of the $H_0$ posterior. The final result thus remain in significant tension with SH0ES: $3.9\sigma$ ($4.5\sigma$) [$4.2\sigma$] for Pantheon+ (DES-Y5) [DES-Dovekie].

{\footnotesize
\bibliographystyle{elsarticle-num}
\bibliography{references}
}

\end{document}